\documentclass[aps,prl,superscriptaddress,preprintnumbers,twocolumn,groupedaddress,showpacs]{revtex4}
\usepackage{txfonts}
\usepackage{graphicx}

\begin{document}
\newcommand{\ppttga}{$pp \to t\bar t\gamma+X~$ }
\newcommand{\ppttg}{$pp \to t\bar t\gamma~$ }
\newcommand{\qqttga}{$q\bar q \to  t\bar t\gamma~$}
\newcommand{\qqttgag}{$q\bar q \to  t\bar t\gamma g~$}
\newcommand{\ggttga}{$gg \to  t\bar t\gamma~$}
\newcommand{\ggttgag}{$gg \to  t\bar t\gamma g~$}

\title{Next-to-leading order QCD corrections to $t\bar t\gamma$ production
at the 7 TeV LHC  }
\author{ Duan Peng-Fei, Zhang Ren-You, Ma Wen-Gan,   \\
Han Liang, Guo Lei, and Wang Shao-Ming  \\
{\small Department of Modern Physics, University of Science and Technology}  \\
{\small of China (USTC), Hefei, Anhui 230026, P.R.China}  }


\begin{abstract}
We present the theoretical predictions including the complete
next-to-leading order (NLO) QCD corrections to the top-quark pair
production in association with a photon at the LHC with
center-of-mass system energy of $7~{\rm TeV}$. The uncertainties of
the leading order (LO) and NLO QCD corrected cross sections due to
the renormalization/factorization scale, and the distributions of
the transverse momenta of final top-quark and photon are studied.
Moreover, we provide the numerical results of the LO, NLO QCD
corrected cross sections and the corresponding $K$-factors with
different photon transverse momentum cuts. We also discuss the
impact of QCD corrections to the \ppttga in case of there existing
the exotic top-quark with an electric charge of $-4e/3$ at the
$7~{\rm TeV}$ LHC.

\end{abstract}

 \maketitle

\par
Since the discovery of top-quark at the Fermilab in
1995\cite{cdftop,d0top}, it has opened up a new research field of
top physics, not only because top-quark breaks the electroweak
symmetry maximally as the heaviest elementary fermion discovered up
to now, but also due to the lack of precise measurement of its
properties. In particular, the couplings of the top-quark to the
electroweak gauge bosons have not yet been precisely measured, which
may turn out to be our first clue of new physics connected with the
electroweak symmetry breaking (EWSB). Moreover, many properties of
the particle are still poorly known, for example, the electric
charge of top-quark, a fundamental quantity characterizing the
electromagnetic interaction between the top-quark and photon, has
not yet been experimentally determined. If we assume that the
electric charge of the top-quark is $q_t =-4e/3$, instead of the
normal value $q_t = 2e/3$, a consistent description of the top
physics could still be obtained\cite{topcharge,consist}. As
suggested by Ref.\cite{ubaur}, studying the production of a
top-quark pair in association with a hard photon is a more effective
way to determine the top-quark charge.

\par
Topics of higher order corrections to the top-quark pair production
in various configurations at hadron colliders are the focus all the
time. As early as late 1980's, the works studying the NLO QCD
corrections to $t\bar t$ production \cite{npds} have been presented.
Recently there has been significant progress in higher order
corrections to $t\bar t$ production\cite{mksm,kjg,kbmz}. NLO QCD
corrections are also known for the associated production processes
involving top-pair production, including $t\bar t H$\cite{wb1},
$t\bar t j$ \cite{sdpu1}, $t\bar t Z$ \cite{atkf} and $t\bar
t\gamma$ \cite{dpf,mss}. Since March 2010 the LHC has begun to run for
the first attempt at $7~{\rm TeV}$. It will continue to operate in
$7~{\rm TeV}$ $pp$ collision mode for some years. Now all the
detectors are recording data, and both the ATLAS and CMS groups are
launching the physics program at the colliding energy of $7~{\rm
TeV}$.

\par
Recently, Prof. Ivor Fleck in ATLAS experimental group suggested us
to extend our NLO QCD calculations for the \ppttga process to the
$7~{\rm TeV}$ LHC, and study the impact of the top-quark electric
charge of $q_t= -4e/3$. He thought that will be helpful for their
experimental data analyses\cite{Fleck}. In this report we calculate
the NLO QCD corrections to the process \ppttga at the $7~{\rm TeV}$
LHC with different photon transverse momentum cuts, and studied
their uncertainties induced by the factorization/renormalization
scale, and the effects to the observables for \ppttga process in
case of there existing the exotic top-quark with an electric charge
of $q_t=-4e/3$. In the paper of Ref.\cite{dpf} we presented the
detail calculations of the NLO QCD corrections to the associated
production of $t\bar{t}\gamma$ at hadron colliders, and presented
the results only for the Tevatron Run II and the $14~{\rm TeV}$ LHC,
but not for the $7~{\rm TeV}$ LHC.

\par
At the leading order (LO), hadronic $t\bar t\gamma$ production
proceeds via $q-\bar{q}$ ($q = u, d, c, s$) annihilation and
gluon-gluon fusion. The NLO QCD corrections to the \ppttga process
contain real and virtual corrections. The Feynman diagrams have been
generated by FeynArts3.4\cite{feyn} automatically, and the Feynman
amplitudes are subsequently reduced by FormCalc5.4\cite{form}
programs. The calculations are carried out by adopting the
dimensional regularization method to regularize the UV and IR
divergences and implying the modified minimal subtraction
($\overline{\rm MS}$) scheme to renormalize the strong coupling
constant and the relevant masses and fields except for the top-quark
and gluon, whose masses and wave functions are renormalized by
applying the on-shell scheme. After renormalization procedure, the
virtual correction is UV-finite. The IR divergences from the
one-loop diagrams will be cancelled by adding the soft real
gluon/light-quark emission corrections by using the two cutoff phase
space slicing method (TCPSS)\cite{twocut}. The remaining collinear
divergences can be absorbed into the parton distribution
functions(PDFs).

\par
We take one-loop and two-loop running $\alpha_{s}$ in the LO and NLO
calculations, respectively\cite{hepdata}, with the number of active
flavors $N_f=5$, the QCD parameters $\Lambda_5^{LO}=165~{\rm MeV}$
and $\Lambda_5^{\overline{MS}}=226~{\rm MeV}$ in the LO and NLO
calculations, respectively. The factorization scale and the
renormalization scale are set to be equal for simplification (i.e.,
$\mu=\mu_f=\mu_r$) and $\mu=\mu_0=m_t$ by default unless otherwise
stated. In the numerical calculations, we take $m_t=171.2~{\rm GeV}$
and $\alpha(m_Z)^{-1}=127.918$\cite{hepdata}. It has been verified
that the total cross section including the NLO QCD corrections is
independent of the cutoffs $\delta_{s}$ and $\delta_{c}$ in adopting
the TCPSS method, which has been shown in Figs.7(a,b) of
Ref.\cite{dpf}. In the following calculations, we fix the soft
cutoff as $\delta_s=10^{-4}$ and collinear cutoff as
$\delta_c=\delta_s/50$. The calculations are carried out at the LHC
with  $pp$ colliding energy $\sqrt{s}=7~{\rm TeV}$. A photon
transverse momentum cut, $p_{T,cut}^{(\gamma)}$, is introduced, and
the radiated photon is required to be relatively hard enough to
fulfill the condition of $p_T^{(\gamma)} > p_{T,cut}^{(\gamma)}$. In
real emission processes, we use the photon isolation cut to separate
photon from the emitted jet\cite{isola}. The events are rejected
unless they fulfills the condition of
\begin{eqnarray}\label{isol}
\sum_i E_{T_i} \theta (\delta - R_{i \gamma}) \leq E_{T_\gamma}
\frac{1-\cos\delta}{1-\cos\delta_0}~~~~~({\rm for~ all}~ \delta \leq
\delta_0),
\end{eqnarray}
where $E_{T_i}$ is the transverse energy of parton $i$,
$E_{T_\gamma}$ is the transverse energy of the photon, $\delta_0$ is
a fixed separation parameter that we set to be $0.4$ in this work,
and $R_{i\gamma}$ is the angular distance between parton $i$ and the
photon defined as
\begin{eqnarray}
R_{i \gamma} = \sqrt { (\eta_i - \eta_\gamma)^2 + (\varphi_i -
 \varphi_\gamma)^2},
\end{eqnarray}
where $\eta$ and $\varphi$ are the pseudorapidity and azimuthal
angle respectively.

\par
\begin{figure}
\centering
\includegraphics[scale=0.7]{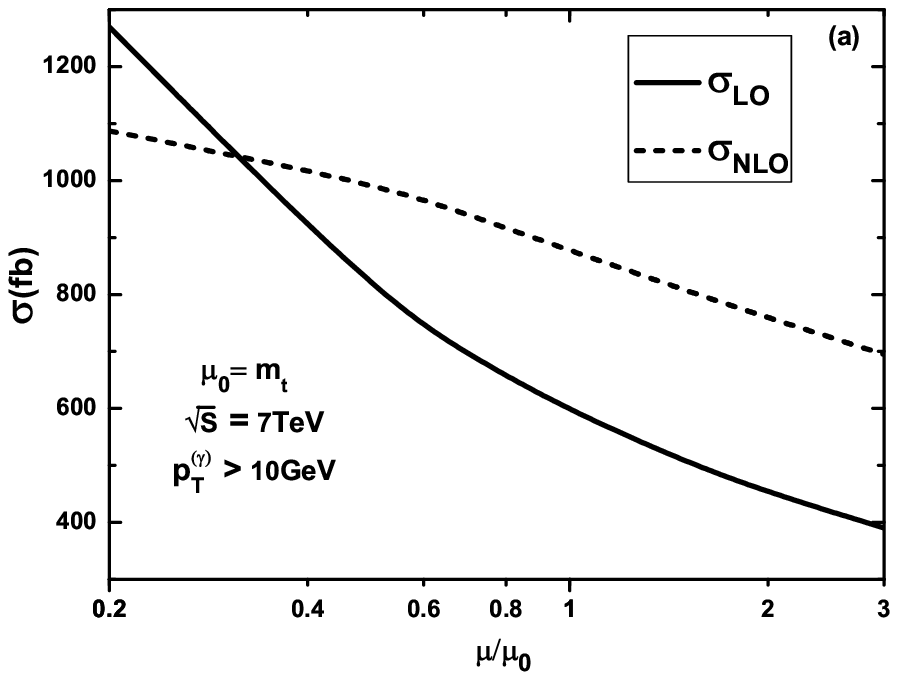}
\includegraphics[scale=0.7]{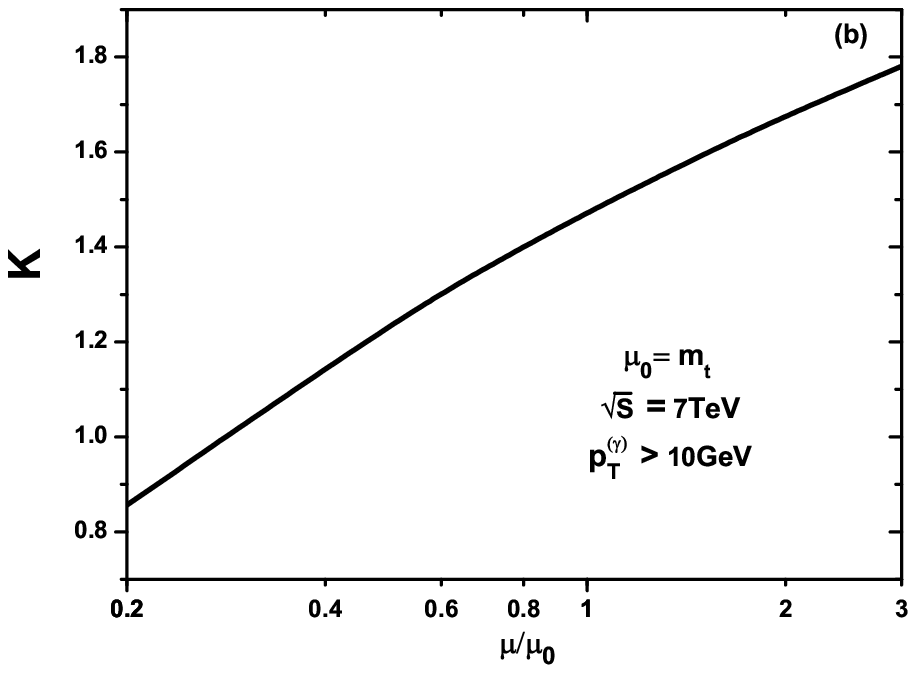}
\caption{\label{fig1} (a) The dependence of the LO and NLO cross
sections on the factorization/renormalization scale at the $7~{\rm
TeV}$ LHC. (b) The corresponding NLO QCD K-factor [$K\equiv
\sigma_{NLO}/\sigma_{LO}$] versus the energy scale. }
\end{figure}

The LO and NLO cross sections for $t\bar t \gamma$ production at the
LHC as the functions of the renormalization and factorization scale
$\mu$ are plotted in Fig.\ref{fig1}(a). The corresponding K-factor
[$K \equiv \sigma_{NLO}/\sigma_{LO}$] is plotted in
Fig.\ref{fig1}(b). In Fig.\ref{fig1}(a) and Fig.\ref{fig1}(b), we
take $p_T^{(\gamma)}>10~{\rm GeV}$, $q_t = 2e/3$, and the colliding
energy in the proton-proton center-of-mass system as $\sqrt s=7~{\rm
TeV}$. As shown in the figures the curve for the NLO cross section
is much less sensitive to $\mu$ than the one for the LO cross
section, which indicates that the NLO QCD correction has obviously
reduced the uncertainty of the cross section on the introduced
parameter $\mu$ in the plotted range of $\mu/\mu_0$.

\par
\begin{figure}
\centering
\includegraphics[scale=0.7]{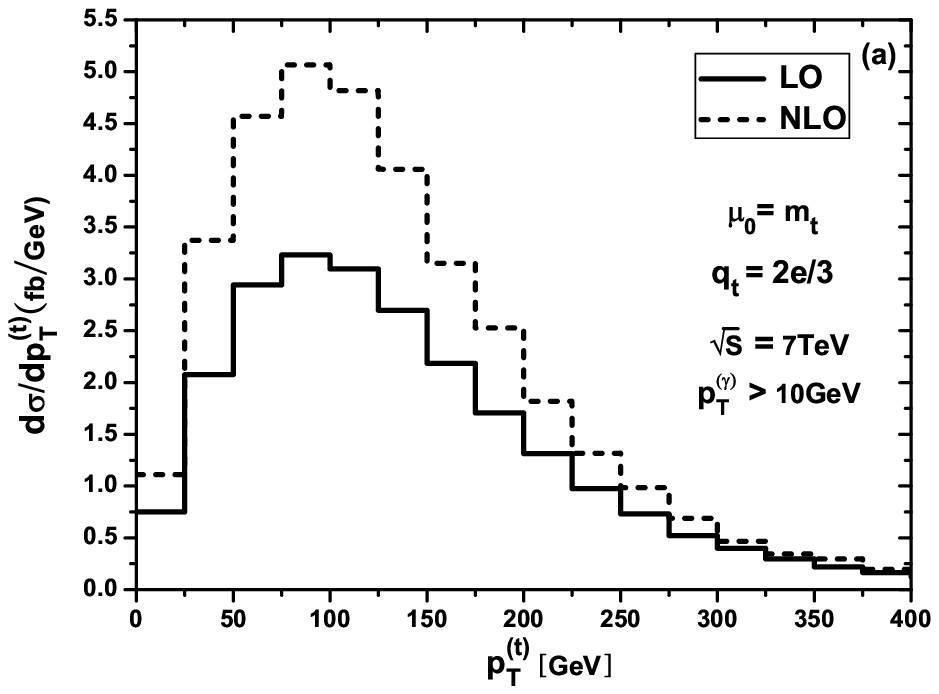}
\includegraphics[scale=0.7]{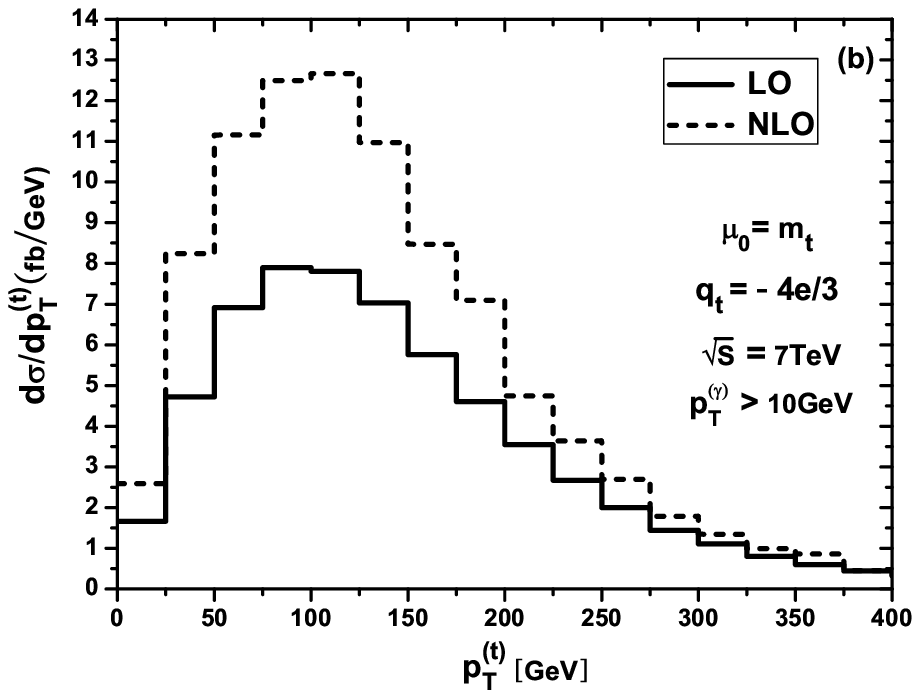}
\caption{\label{fig2} (a) The LO and NLO distributions of the
transverse momentum of the normal top-quark with $q_t=2e/3$. (b) the
LO and NLO distributions of the transverse momentum of the exotic
top-quark with $q_t=-4e/3$, $\mu=m_t$, $p_T^{(\gamma)}>10~{\rm GeV}$
and $\sqrt{s}=7~{\rm TeV}$. }
\end{figure}
\begin{figure}
\centering
\includegraphics[scale=0.7]{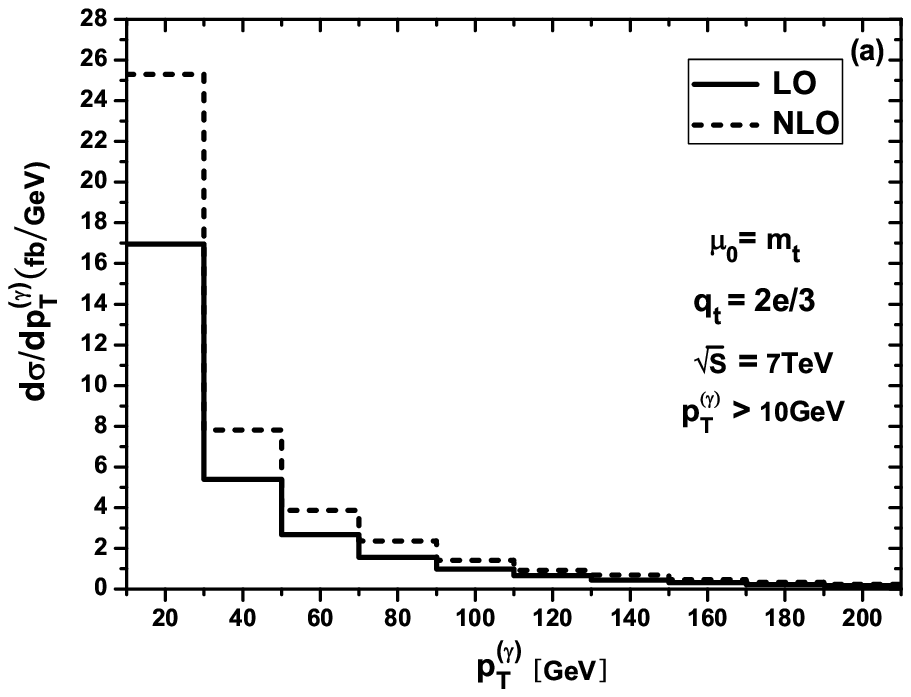}
\includegraphics[scale=0.7]{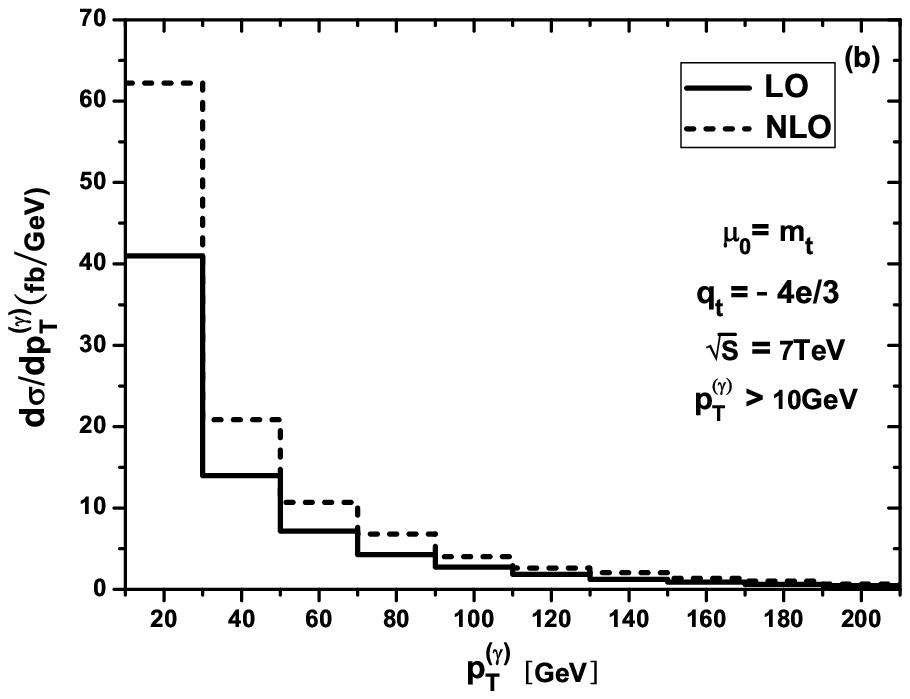}
\caption{\label{fig3} The LO and NLO distributions of the transverse
momentum of the photon at LHC with $\mu=m_t$,
$p_T^{(\gamma)}>10~{\rm GeV}$ and $\sqrt{s}=7~{\rm TeV}$. (a)
top-quark with electric charge $q_t=2e/3$. (b) top-quark with
electric charge $q_t=-4e/3$. }
\end{figure}

The LO and NLO distributions of the transverse momenta for the normal
top-quark (with electric charge $2e/3$) and exotic top-quark (with
electric charge $-4e/3$) at the LHC, are depicted in
Fig.\ref{fig2}(a) and Fig.\ref{fig2}(b), respectively. The LO and
NLO distributions of the transverse momenta for the photon with
different top-quark electric charge ($2e/3$ or $-4e/3$) are depicted
in Fig.\ref{fig3}(a) and Fig.\ref{fig3}(b), respectively.
Figs.\ref{fig2} and \ref{fig3} demonstrate that the NLO QCD
corrections enhance significantly the distributions of $p_T^{(t)}$
and $p_T^{(\gamma)}$ for both the normal top-quark and the exotic
top-quark. From the distributions of $p_T^{(\gamma)}$ in both
Fig.\ref{fig3}(a) and Fig.\ref{fig3}(b), we conclude that most of
the photons in the events of \ppttga are produced in low transverse
momentum range at the LHC.

\par
Now we study the impacts of the photon transverse momentum cut
$p_{T,cut}^{(\gamma)}$ and the exiotic top-quark electric charge
$q_t$ on the cross section and K-factor. We calculate the LO and NLO
cross sections and the corresponding K-factor [$ K\equiv
\sigma_{NLO}/\sigma_{LO}$] with $\mu = m_t$, $\sqrt{s}=7~{\rm TeV}$,
$p_T^{(\gamma)}>10~{\rm GeV}$ (or $15~{\rm GeV}$) and $q_t = 2e/3$
(or $-4e/3$). The numerical results are presented in Table
\ref{tab1}. Our numerical results turn out that the K-factor varies
only slightly with the variation of either the photon transverse
momentum cut or the top-quark electric charge (about $3\%$ for the
latter), although the LO and NLO cross sections strongly depend on
the photon transverse momentum cut and the top-quark electric
charge. We know that if the photon is merely radiated from the
top-quarks, the cross section for \ppttga production would be
proportional to the square of the top-quark electric charge, which
implies that the cross section with top-quark electric charge
$q_t=-4e/3$ would be four times of that with top-quark electric
charge $q_t=2e/3$. Actually, both the LO and the NLO Feynman graphs
include the diagrams with photon radiating from light-quark
respectively, which makes the total cross sections in the exotic
top-quark case less than three times of that in the normal top-quark
case.
\begin{table}
\begin{center}
\begin{tabular}{|c|c|c|c|c|}
\hline   $ $   & $\sigma_{LO}$(fb)  &$\sigma_{NLO}$(fb) &$K$  \\
\hline   $p_T^{(\gamma)}>10~{\rm GeV}$, $q_t = 2e/3$  & 596.0(2)  & 878(2)   & 1.4732 \\
\hline   $p_T^{(\gamma)}>10~{\rm GeV}$, $q_t = -4e/3$ & 1513(1)   & 2276(5)  & 1.5043 \\
\hline   $p_T^{(\gamma)}>15~{\rm GeV}$, $q_t = 2e/3$  & 454.6(2)  & 668(2)   & 1.4694 \\
\hline   $p_T^{(\gamma)}>15~{\rm GeV}$, $q_t = -4e/3$ & 1179(1)   & 1775(5)  & 1.5055\\
\hline
\end{tabular}
\end{center}
\begin{center}
\begin{minipage}{7cm}
\caption{\label{tab1} The LO and NLO cross sections and the
corresponding NLO QCD K-factor [$K\equiv \sigma_{NLO}/\sigma_{LO}$]
with $\mu = m_t$, $\sqrt{s}=7~{\rm TeV}$ and different values of
photon transverse momentum cut $p_{T,cut}^{(\gamma)}$ and top-quark
electric charge $q_t$. }
\end{minipage}
\end{center}
\end{table}

\par
In summary, the calculations for the top-pair production associated
with a photon at the $7~{\rm TeV}$ LHC have been performed with the
complete NLO QCD corrections. We investigate the dependence of the
LO and NLO QCD corrected integrated cross sections on the
factorization/renormalization energy scale. We present predictions
for the LO and NLO differential and integrated cross sections at the
$7~{\rm TeV}$ LHC in the cases of top-quark with $q_t=2e/3$ and
$-4e/3$. We demonstrate that the uncertainty of the LO cross section
caused by the introduced energy scale $\mu$ is significantly
improved by including the NLO QCD corrections. We study also the
effects of different values of photon transverse momentum cut and
top-quark charge on the $K$-factor. Our numerical results show that
the K-factor is not very sensitive to both the transverse momentum
cut of photon ($p_{T,cut}^{(\gamma)}$) and the top-quark electric
charge ($q_t$). The K-factor variation is about $3\%$, when we
change $q_t$ from $2e/3$ to $-4e/3$. But the $p_{T,cut}^{(\gamma)}$
and $q_t$ impact the total integrated cross sections obviously.

\vskip 10mm
\par
\noindent{\large\bf Acknowledgments:} This work was supported in
part by the National Natural Science Foundation of China
(No.10875112, No.11075150, No.11005101), and the Specialized
Research Fund for the Doctoral Program of Higher Education
(No.20093402110030).


\vskip 10mm
\begin{thebibliography}{99}
\bibitem{cdftop}
  Abe F, {\it et al.} (CDF Collaboration) 1995 Phys. Rev. Lett. {\bf 74} 2626

\bibitem{d0top}
  Abachi S, {\it et al.} (D\O\ Collaboration) 1995 Phys. Rev. Lett. {\bf 74} 2632

\bibitem{topcharge}
  Chang D, Chang W F and Ma E 1998 Phys. Rev. {\bf D59} 091503

\bibitem{consist}
  Abazov V M, {\it et al.} (D0 Collaboration) 2007 Phys. Rev. Lett. {\bf 98} 041801

\bibitem{ubaur}
  Baur U, Buice M and Orr L H 2001 Phys. Rev. {\bf D64} 094019

\bibitem{npds}
  Nason P, Dawson S and Ellis R K 1988 Nucl. Phys. {\bf B303} 607; 
  Beenakker W, Kuijf H, van Neerven W L and Smith J 1989 Phys. Rev. {\bf D40} 54

\bibitem{mksm}
  Melnikov K and Schulze M 2009 JHEP {\bf 08} 049; 
  Czakon M and Mitov A, Nucl. Phys. {\bf B824} (2010) 111 

\bibitem{kjg}
  Korner J G, Merebashvili Z and Rogal M 2008 Phys. Rev. {\bf D77} 094011;
  Czakon M 2008 Phys. Lett. {\bf B664} 307; 
  Bonciani R, Ferroglia A, Gehrmann T, Maitre D and Studerus C 2008 JHEP {\bf 07} 129; 
  Anastasiou C and Aybat S M 2008 Phys. Rev. {\bf D78} 114006

\bibitem{kbmz}
  Bonciani R, Ferroglia A, Gehrmann T and Studerus C 2009 JHEP {\bf 08} 067; 
  Beneke M, Falgari P and Schwinn C 2010 Nucl. Phys. {\bf B828} 69; 
  Czakon M, Mitov A and Sterman G F 2009 Phys. Rev. {\bf D80} 074017

\bibitem{wb1}
 Beenakker W, {\it et al.} 2001 Phys. Rev. Lett. {\bf 87} 201805;
 Reina L and Dawson S 2001 Phys. Rev. Lett. {\bf 87} 201804

\bibitem{sdpu1}
 Dittmaier S, Uwer P and Weinzierl S 2007 Phys. Rev. Lett. {\bf 98} 262002;
 Melnikov K and Schulze M 2010 Nucl. Phys. {\bf B840} 129 

\bibitem{atkf}
 Lazopoulos A, McElmurry T, Melnikov K and Petriello F 2008 Phys. Lett. {\bf B666} 62 

\bibitem{dpf}
 Duan P F, Ma W G, Zhang R Y, Han L, Guo L and Wang S M 2009 Phys. Rev.{\bf D80}, 014022

\bibitem{mss}
 Melnikov K, Scharf A and Schulze M Phys. Rev.{\bf D83} 074013

\bibitem{Fleck}
 Private communications between Fleck I and authors

\bibitem{feyn}
 Hahn T 2001 Comput. Phys. Commun. {\bf 140} 418

\bibitem{form}
 Hahn T and Perez-Victoria M 1999 Comput. Phys. Commun. {\bf 118} 153

\bibitem{twocut}
  Harris B W and Owens J F 2002 Phys. Rev. {\bf D65} 094032

\bibitem{hepdata}
  Amsler C, {\it et al.} 2008 Phys. Lett. {\bf B667} 1 

\bibitem{isola}
 Frixione S 1998 Phys. Lett. {\bf B429} 369 



\end{thebibliography}
\end{document}